\documentclass[a4paper,11pt]{article}
\usepackage{pos}
\usepackage{graphicx} 
\usepackage{epsfig}
\usepackage{epstopdf}
\usepackage{amsfonts}
\usepackage{amsmath}
\makeatletter
\let\c@lofdepth\relax
\let\c@lotdepth\relax
\makeatother
\usepackage{subfigure}
\usepackage{tabularx}

\title{Model selection results from different BAO datasets -- DE models and $\Omega_K$CDM }
\ShortTitle{Model selection from marginalised BAO datasets}

\author*[a]{Denitsa Staicova}

\affiliation[a]{
$^{1}\,$ Institute for Nuclear Research and Nuclear Energy, Bulgarian Academy of Sciences, Sofia, Bulgaria}

\emailAdd{dstaicova@inrne.bas.bg}

\abstract{The use of the baryonic acoustic oscillations (BAO) datasets offers a unique opportunity to connect the early universe and the late one. In this proceeding, we discuss recent results that used a marginalised likelihood to remove the $H_0-r_d $ degeneracy and then tested it on different dark energy (DE) models. It was found that this approach which does not rely on calibration on $r_d$ or $H_0$, allows us to obtain results, comparable to the ones calculated with standard likelihoods. Here we emphasize on the major differences that we observed for the two different BAO datasets that we employed -- a transversal one, containing only angular BAO measurements, and a mixed one, containing both angular and radial BAO measurements. We see that the two datasets have different statistical preferences for DE models and also different preference for the curvature of the universe.}

\FullConference{%
  Corfu Summer Institute 2022 "School and Workshops on Elementary Particle Physics and Gravity",\\
  28 August - 1 October, 2022\\
  Corfu, Greece}

\date{February 2023}

\begin{document}

\maketitle

\section{Introduction}
The degeneracy between the Hubble parameter $H_0$ and the sound horizon horizon $r_d$ is a known problem, sometimes called the $H_0-r_d$ tension (\cite{Knox:2019rjx, Abdalla:2022yfr}). For the concordance $\Lambda$CDM cosmology, the sound horizon, which makes the baryonic acoustic oscillations (BAO) a standard ruler, is known and it depends on the ratio of baryonic to radiative content of the early universe. The problem, however, is that BAO measurements, both radial and transversal, always measure the quantity $H_0 r_d$. For this reason, in order to disentangle $H_0$ and $r_d$, it is required  either a prior knowledge coming from the early universes (i.e a choice of $r_d$) \cite{Aghanim:2018eyx}, or a prior knowledge coming from the late universe (i.e. a prior on $H_0$) \cite{Riess:2021jrx, Riess:2022mme}. In both cases, choosing a prior on either quantity will bring certain assumptions in our model, which will effectively lead to a calibration of the BAO measurement with either the early universe or the late one. While local measurements of $H_0$  \cite{Riess:2021jrx, Riess:2022mme} are considered model-independent and are done with ever increasing precision, thanks to the newer instruments such as JWST \cite{Yuan:2022edy} and the improvement of the methods, they are still at odds with the early universe measurements from the Planck mission \cite{Aghanim:2018eyx}. The discrepancy between local universe and the early universe has been seen in different quantities - the Hubble constant $H_0$, the $\sigma_8-S_8$ and other anomalies \cite{Abdalla:2022yfr}. The tensions in cosmology have challenged our models for years now and have led to a lot of works looking for ways to resolve the problem or at least to mitigate it (for recent reviews, see \cite{Abdalla:2022yfr, Poulin:2023lkg}).    

In a series of articles \cite{Benisty:2020otr, Benisty:2022psx, Staicova:2021ntm, Staicova:2022zuh}, we considered the applications of different BAO datasets to constrain cosmological parameters and we looked for a way around the mentioned calibration leading to the $H_0-r_d$ tension. Here we will discuss the approach in \cite{Staicova:2021ntm}, in which we used a marginalisation procedure to integrate out of the likelihood $\chi^2$ the factor $H_0 r_d$. This procedure allows us to completely avoid setting any prior on $H_0$ and $r_d$ because our $\chi^2$ no longer depends on them. Then, we use this uncalibrated by early or late universe likelihood, to perform full Bayesian analysis for different models of dark energy (DE). We find that without any other assumptions, the BAO datasets cannot constrain well the DE models, but by adding the type IA supernova dataset, the errors on the DE parameters become smaller. We also find that the two datasets we use have statistical preference for different cosmological models, which we will discuss in detail.

\section{Constraining Dark Energy models}

\subsection{The equation of state of the Universe}
We assume a Friedmann-Lema\^itre-Robertson-Walker metric, with a standard Friedman equation $(H(z)/H_0)= E(z)$ for : 
\begin{equation}
      E(z)^2= \Omega_{m} (1+z)^3 + \Omega_{K} (1+z)^2 + \Omega_{\Lambda}(z),
    \label{eq:hz}
\end{equation}
 where $z$ is the redshift and  the scale factor is $a = 1/(1+z)$, $H(z) := \dot{a}/a$ is the Hubble parameter at redshift $z$ and $H_0$ is the Hubble parameter today. $\Omega_{m}$, $\Omega_{\Lambda}$, and $\Omega_{K}$ are the fractional densities of matter, DE, and the spatial curvature at redshift $z=0$. 

We consider two types of dark energy models. First, we consider an expansion of $\Lambda$CDM in the form of the Chevallier-Polanski-Linder parametrization (CPL) (\cite{Chevallier:2000qy,Linder:2002et,Linder:2005ne,Barger:2005sb}) of the $wwaCDM$ model:
\begin{equation}
\Omega_{\Lambda} \left(z\right) = \Omega_{\Lambda}^{(0)}  \exp\left[\int_0^{z} \frac{3(1+w(z')) dz'}{1+z'}\right]
\label{eq:integOmeLambda}
\end{equation}
in which we considered three possible models:
\begin{equation}
w(z) =\begin{cases} w_0 + w_a z & \text{Linear }  \\ 
 w_0 + w_a \frac{z}{z+1} & \text{CPL }  \\ 
w_0 - w_a \log{(z+1)} & \text{Log }  \end{cases}, 
\end{equation}
which recover the $\Lambda$CDM for $w_0=-1, w_a=0$.

As an alternative to $\Lambda$CDM, we consider the phenomenologically Emergent Dark Energy (pEDE) model \cite{Li:2019yem, Li:2020ybr} and its generalization (gEDE). gEDE is described by: 
\begin{equation}
\Omega_{DE}(z)=\Omega_\Lambda \frac{1-\tanh(\bar{\Delta} \log_{10}(\frac{1+z}{1+z_t}))}{1+\tanh(\bar{\Delta} \log_{10}({1+z_t})},
\end{equation}
with pEDE-CDM recovered for $\bar{\Delta}=1$, and $\Lambda CDM$ for $\bar{\Delta}=0$. Here, the transitional redshift $z_t$ is found from ${\Omega}_{DE}(z_t)=\Omega_m (1+z_t)^3$, see \cite{Li:2020ybr}.

 {The radial BAO projection $D_H(z)= c/H(z)$ is found from:}
\begin{equation}
\frac{D_H}{r_d} = \frac{c}{H_0 r_d} \frac{1}{E(z)}.
\end{equation}
 { The tangential BAO measurements expressed in terms of the angular diameter distance $D_\textrm{A}/r_d$ are:}
\begin{subequations}
\begin{equation}
\frac{D_A}{r_d} = \frac{c}{H_0 r_d} f(z),
\end{equation}
where:
\begin{equation}
f\left(z\right) = \frac{1}{(1+z)  \sqrt{|\Omega_{K}|}  } \textrm{sinn}\left[|\Omega_{K}|^{1/2}\Gamma(z)\right]. 
\end{equation}
\end{subequations}
 
The BAO angular scale measurement $\theta_{BAO}(z)$ needed for the $\theta_{BAO}$ dataset is : 
\begin{equation}
\theta_{BAO}\left(z\right) = \frac{r_d}{\left(1+z\right)D_A(z)} = \frac{H_0 r_d}{c}h(z),
\end{equation}
with:

 {\begin{equation}
h\left(z\right) = \frac{1}{ (1+z) f(z)}
\end{equation}}

The SNIa measurements are described by their distance modulus which is related to the luminosity distance ($D_A=d_L(z)/(1+z)^2$) through:
\begin{equation}
        \mu_B (z) - M_B = 5 \log_{10} \left[ d_L(z)\right] + 25  \,,
\label{eq:dist_mod_def}
\end{equation}
where $d_L$ is measured in units of Mpc, and $M_B$ is the absolute magnitude.

\subsection{The $\chi^2$ redefinition}
  For the BAO points,  we redefine $\chi^2$ in a way that isolates $\frac{c}{H_0 r_d}$, i.e. we perform a maginalization procedure\cite{Lazkoz:2005sp,Basilakos:2016nyg,Anagnostopoulos:2017iao,Camarena:2021jlr}. Omitting the details that can be found in \cite{Staicova:2021ntm}, the final $\chi^2$ becomes:
  \begin{equation}
\tilde{\chi}^2_{BAO} = C-\frac{B^2}{A} + \log\left(\frac{A}{2 \pi}\right).
\label{eq:chi2BAO}
\end{equation} 
where:
\begin{subequations}
\begin{equation}
A  = f^j(z_i) C_{ij} f^i(z_i),
\end{equation}
\begin{equation}
B = \frac{f^j(z_i) C_{ij} v_{model}^i(z_i) + v_{model}^j(z_i) C_{ij} f^i(z_i)}{2},
\end{equation}
\begin{equation}
C = v_j^{model} C_{ij} v_i^{model}.
\end{equation}
\label{eq:termChi2}
\end{subequations}

Here $\vec{v}_{obs}$ is the vector of the observed points  {at each $z$ (i.e., $D_M/r_d$, $D_H/r_d$, $D_A/r_d$ or $\theta_{BAO}$)} and $\vec{v}_{model}$ is the theoretical prediction of the model. $C_{ij}$ is the covariance matrix (for uncorrelated points, it becomes a diagonal matrix with elements equal to the inverse errors $\sigma_i^{-2}$.

For the $\theta_{BAO}$ data, the $\chi^2$ has the same form, but the coefficients are as follows: 
\begin{subequations}
\begin{equation}
A_{\theta}  = \sum_{i = 1}^{N} \frac{h(z_i)^2}{\sigma_i^2},
\end{equation}
\begin{equation}
B_{\theta}  = \sum_{i = 1}^{N} \frac{{\theta}_{D}^{i}\, h(z_i) }{\sigma_i^2},
\end{equation}
\begin{equation}
C_{\theta}  = \sum_{i = 1}^{N} \frac{\left({\theta}_{D}^{i}\right)^2}{\sigma_i^2}.
\end{equation}
\end{subequations}

For the $SN$ data, we assumed no prior constraint on $M_B$, and we integrated the probabilities over $M_B$ and $H_0$ \cite{DiPietro:2002cz,Nesseris:2004wj,Perivolaropoulos:2004yr,Lazkoz:2005sp} to get the marginalized $\chi^2$:
\begin{equation}
\tilde{\chi}^2_{SN} = D-\frac{E^2}{F} + \ln\frac{F}{2\pi},
\end{equation}
where:
\begin{subequations}
\begin{equation}
D = \sum_i \left( \Delta\mu \, C^{-1}_{cov} \, \Delta\mu^T \right)^2,
\end{equation}
\begin{equation}
E = \sum_i \left( \Delta\mu \, C^{-1}_{cov} \, E \right),
\end{equation}
\begin{equation}
F = \sum_i  C^{-1}_{cov}  ,
\end{equation}
\end{subequations}
where $\Delta\mu =\mu_{}^{i} - 5 \log_{10}\left[d_L(z_i)\right)$, $E$ is the unit matrix, and $C^{-1}_{cov}$ is the inverse covariance matrix of the dataset \cite{Deng:2018jrp}. 

\subsection{Datasets and methods}
We use two different BAO datasets, to which we add the binned Pantheon supernovae dataset with its covariance matrix. The BAO datasets can be found summarized in \cite{Staicova:2021ntm}. First, we use a BAO dataset, denoted as $BAO$,  which contains various angular data-points combined with few points from the most recent to date eBOSS data release (DR16), which come as  angular ($D_M$) and radial ($D_H$) measurements and their covariance. The second dataset is denoted as $BAO{_\theta}$ and it consists of 15 transversal BAO measurements used for clustering analysis \cite{Nunes:2020hzy}. These points have the advantage that they are uncorrelated and that they do not assume a fiducial cosmology, particularly on the $\Omega_K$ parameter, which is included in the standard BAO analysis \cite{Nunes:2020hzy}. To the BAO points, we add the binned Pantheon dataset, which contains $40$ supernovae luminosity measurements in the redshift range $z\in (0.01,2.3)$ \cite{Pan-STARRS1:2017jku} (called here "SN"). 

The priors we used can be found in \cite{Staicova:2021ntm}. We use the MCMC nested sampler, implemented in the open-source package {\textit{Polychord}} \cite{Handley:2015fda} with the {\textit{GetDist}} package \cite{Lewis:2019xzd} to present the results.

\subsection{Results}

After performing the MCMC, the results we obtained can be summarized as follows. The BAO-only datasets do not constrain the DE models parameters, particularly the parameter $w_a$ is practically unconstrained and $w_0$ is found with a big error. The only parameter that gets a good constraint is $\Omega_m$. The combined BAO + SN dataset reduces the errors significantly and allows to put better constraints on $w_0$ but again, it does not constrain well $w_a$. The results we get for the DE models for the combined datasets are in the table below:

\begin{table*}[!h]
    \centering
    \begin{tabular}{c|c|c}
\hline
        Model & $w_0$ & $w_a$\\
        \hline
         \multicolumn{3}{c}{BAO +SN}\\ 
         \hline
         wCDM &  $ -0.99\pm 0.05$ & \\
         wwaCDM & $ -1.18\pm 0.14$ &  $-0.38\pm 0.67$\\ 
         \hline
         \multicolumn{3}{c}{$BAO{_\theta}+SN$}\\ 
          \hline
           wCDM & $ -1.08\pm 0.14$ & \\
           wwaCDM & $-1.09\pm 0.09$ & $ -0.31\pm 0.74$ \\
           \hline
    \end{tabular}
    \label{tab:my_label}
\end{table*}

 This result is consistent  with the SDSS-IV results \cite{eBOSS:2020yzd}, which obtains: $w_0 = -0.939 \pm 0.073$, $w_a= -0.31\pm 0.3$ for the combination BAO+SN+CMB, but $w_0= -0.69 \pm 0.15$ when only the BAO dataset is used. We see that our result is comparable to this, even though some precision is lost due to the marginalization procedure. 

A very interesting result we obtained is that the two BAO datasets do not prefer the same DE models. This becomes very obvious when one uses statistical measures to compare the DE models we consider to the $\Lambda$CDM model. In \cite{Staicova:2021ntm} we use 4 different measures: the Akaike information criterion (AIC), the Bayesian information criterion (BIC), the deviance information 
criterion (DIC), and the Bayes factor (BF) \cite{Liddle:2007fy}. The AIC and BIC favor systematically the $\Lambda$CDM model as the model with the least number of parameters, with the only exclusion being the pEDE model. For this reason, in this work, we present only the other two measures DIC and BF.  

The DIC estimator is defined as:
\begin{equation}
{\rm DIC} =  2\overline{(D(\theta))}-D(\overline{\theta}), 
\end{equation}
where $\theta$ is the vector of parameters being varied in the model, the overline denotes the usual mean value, and $D(\theta) = -2\ln(\mathcal{L(\theta)})+C$, where $C$ is a constant. With this definition, we calculate the difference $\Delta \text{DIC}_{\text{model}}=\text{DIC}_{\text{$\Lambda$CDM}}-\text{DIC}_{\text{model}}$ . A positive $\Delta$DIC points to a preference toward the DE model, negative -- toward $\Lambda$CDM with $|\Delta\text{DIC}|\geq 2$ signifying a possible tension. 

The Bayes factor is defined as:
$$B_{ij}= \frac{p(d|M_i)}{p(d|M_j)},$$
where $p(d|M_i)$ is the Bayesian evidence for model $M_i$, which we calculate numerically with  Polychord. We use the $ln(B_{0i}),$ where "0" is $\Lambda$CDM, which we compare with all the other models (denoted by the index "i") . According to the Jeffry's scale, $ln(B_{ij}) <1$ is inconclusive for any of the models, for 1-2.5 one finds weak support for the model "i" and above 2.5, there is moderate and strong support for the model "i". A minus sign gives the same for model "j" \cite{Jeffreys:1939xee}.

\begin{figure}[!h]
        \centering
\includegraphics[width=0.45\textwidth]{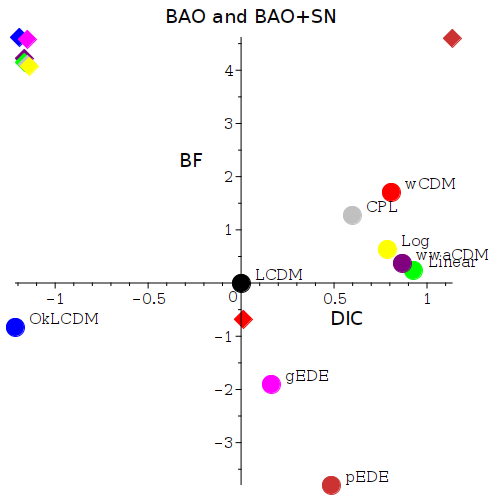}
\includegraphics[width=0.45\textwidth]{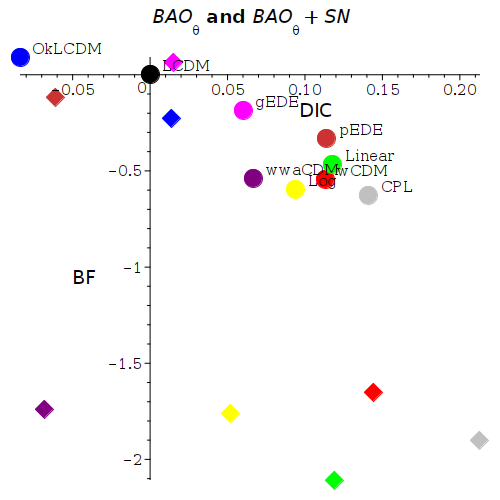}
\caption{\it{The DIC-BF plane for the BAO and BAO + SN points (left) and $BAO_\theta$ and $BAO_\theta+SN$ points (right). We present everywhere the BAO only points with solid spheres, the BAO+SN points with diamonds.}}
        \label{fig:models}
\end{figure}

The results from Table 3 and 4 in \cite{Staicova:2021ntm} are presented on Fig. \ref{fig:models}, where the solid circle signifies the BAO dataset and the diamond, the BAO + SN datset.  On the left figure, the values of the BAO + SN points are divided by 4 in both directions to fit the plot. In order for a model to be better than $\Lambda$CDM in DIC, it has to be in the right side of the plot, in order for it to be better with respect to BF, it has to be in the lower part of the plot. So the best models to challenge $\Lambda$CDM are the ones in the lower right corner of the plots. We see that for the BAO points, these models are only pEDE and gEDE. For the BAO+SN points, there is no such model except maybe for wCDM. On the other hand, the $BA0_\theta$ points exhibit a totally different behavior. We see that there are a lot of comparable to $\Lambda$CDM model for the $BAO_\theta$ alone points, but there are also such points for the $BAO_\theta+SN$ - basically all of the CPL models are in the right part of the plot, even though with respect to DIC, the distance is very small, meaning a not significant preference for DE models.  The BF value, however, signifies some tension with $\Lambda$CDM. One can see that this dataset seems to be much more favorable to the different DE models from statistical point of view. The errors for both datasets are comparable. Such results have also been found in other model-independent approaches \cite{Bernardo:2022pyz, Mehrabi:2022mnr}. 

Another important difference between the two BAO datasets can be seen is in the $\Omega_K$CDM model. From our results presented in \cite{Staicova:2021ntm} we see that the$BAO+SN$ dataset prefers a closed universe ($\Omega_K=-0.21\pm0.07$), while the $BAO{_\theta}+SN$ dataset prefers a flat one ($\Omega_K=-0.09\pm0.15$). Such strong deviation from the flat universe seems questionable, and because of this, we investigate it further  below. We study the dependency of the results on the prior on $\Omega_K$ using two priors -- the standard one $\Omega_K\in [-0.3,0.3]$ and a "large" one -- $\Omega_K \in [-0.7,3]$. One can see the results of the different priors on Fig. \ref{fig:BAO_Ok}. The full posteriors are in the Appendix.

\begin{figure}[!h]
        \centering
\includegraphics[width=0.45\textwidth]{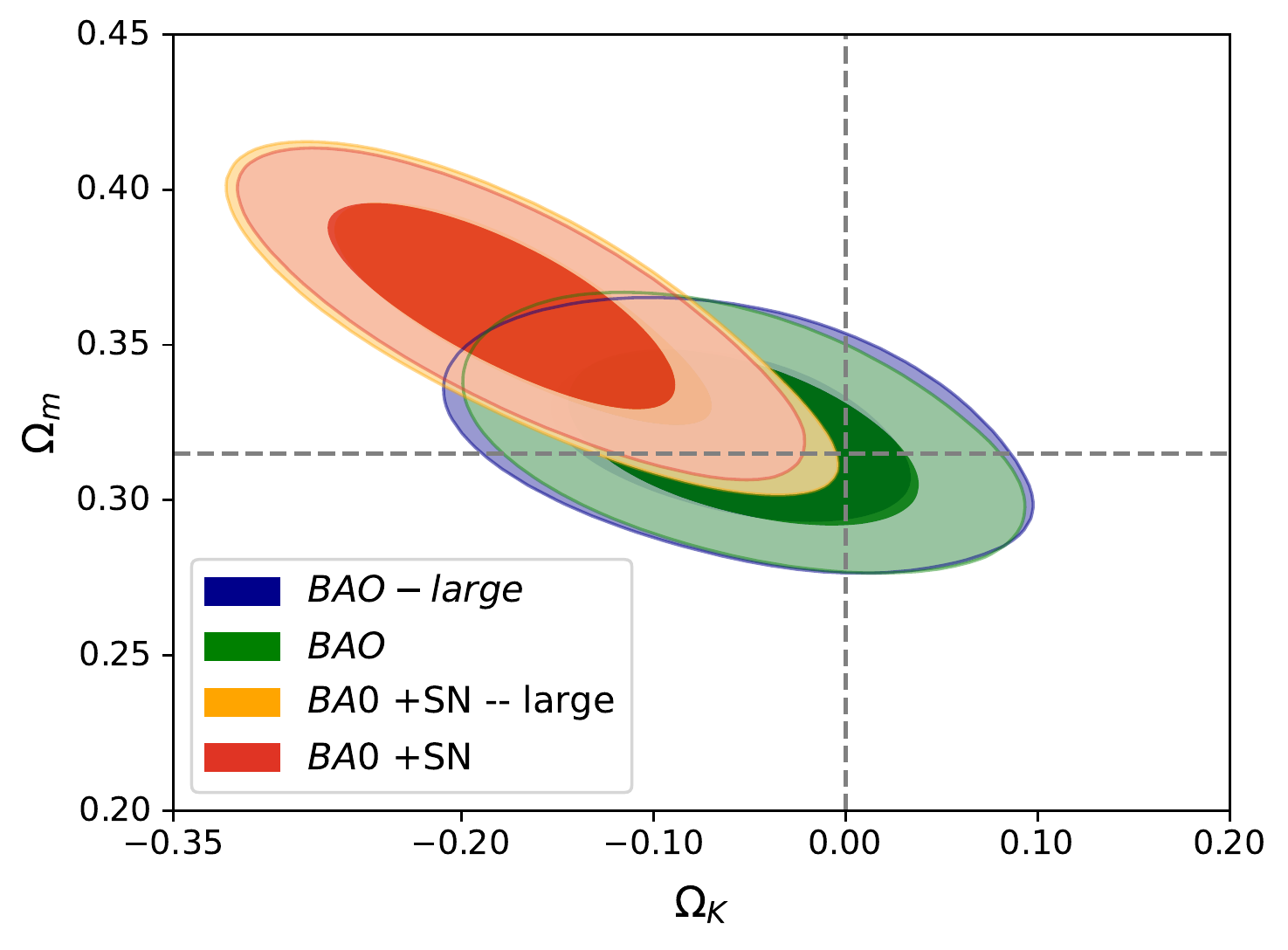}
\includegraphics[width=0.45\textwidth]{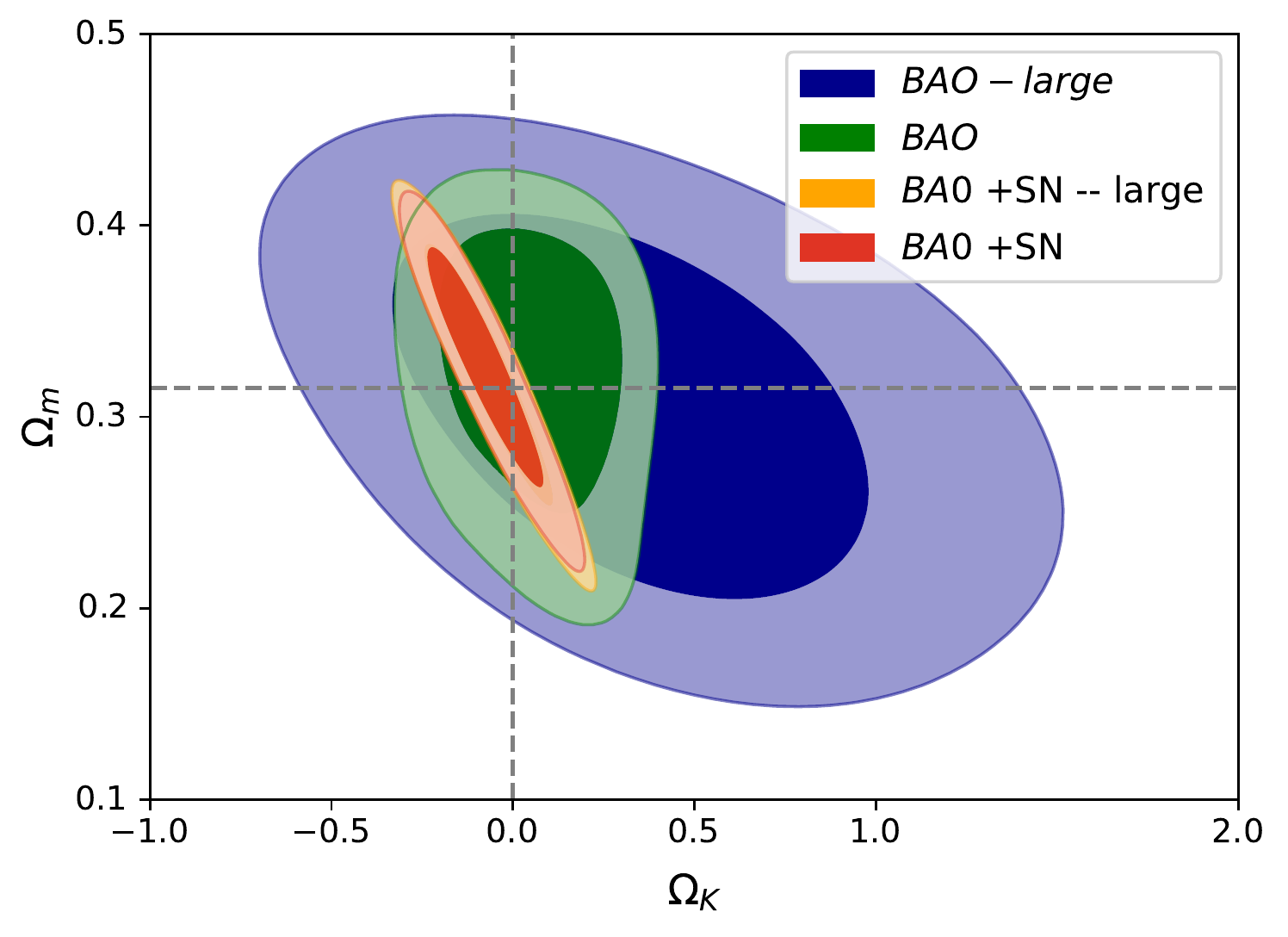}
\caption{\it{Posterior distribution for $\Omega_m$ and $\Omega_K$ for BAO and $BAO_\theta$ datasets to the left and right, respectively}}
        \label{fig:BAO_Ok}
\end{figure}

From the figure, one can see that there is a significant difference with respect to the constraints on $\Omega_K$ and $\Omega_m$ for the two datasets. The $BAO_\theta$ points, with and without the addition of the SN points are centered around $0$ and the $BAO_\theta$ points alone basically do not constrain $\Omega_K$ at all. Notably, in the negative direction, there is a problem with the integration, thus the prior is smaller, so the best constraint we find is $|\Omega_K|<1$. For the BAO points, on the other hand, we see that the mean value is centered at negative $\Omega_K$ and adding the SN points do not help. One can consider two possible explanations of this result. Either it is due to the fact that we lose sensitivity to the $\Omega_K$ parameter in this marginalised form of the $\chi^2$, especially for the SN points. Or this is a result of the specific processing of the $BAO_\theta$ dataset that cleaned up any fiducial cosmology, particularly with respect to the spatial curvature, rendering it insensitive to this parameter.  Or maybe we are seeing both effects simultaneously, with the marginalisation removing an important leverage on the value of $\Omega_K$.  Note, on the figure we have denoted the flat cosmology with a line and also, the Planck matter density $\Omega_m$. One can see that in the BAO+SN case, the negative curvature is achieved on the price of higher $\Omega_m$, which leads also to higher $\Omega_\Lambda$, since due to the marginalization procedure, there is no other parameter to compensate for the non-zero $\Omega_K$.

The discussion on the value of $\Omega_K$ is not new \cite{Ryan:2018aif, Vagnozzi:2020dfn, DiValentino:2019qzk, Vagnozzi:2020rcz, Abdalla:2022yfr}. Recently, \cite{Yang:2022kho} studied different DE cosmologies and found that there are indications of negative $\Omega_K$ in most of them. An interesting study from the point of view of renormalization group approach argues that the flat universe is the only one offering scale-free, non-singular background for
cosmological perturbations\cite{Jimenez:2022asc}. From the literature we see that while the deviation from the flat universe is usually small, model-independent approaches seem to give larger deviations, similar to the ones we obtained, for example \cite{Wei:2016xti, Ryan:2018aif, Stevens:2022evv}. Also while strange that adding the SN data leads to bigger $\Omega_K$, it can be seen also in \cite{Yang:2022kho} where CMB+BAO has bigger $\Omega_K$ than CMB+Pantheon. The Planck data alone also seem to prefer a closed universe \cite{Aghanim:2018eyx}. 

\subsection{Discussion}
We have summarized our results on the use of a $\chi^2$ marginalization procedure on a BAO + SN datasets that we utilized to study different DE modes. The results on the DE models are very similar to the already published in the literature, which shows that our procedure can be useful for studying alternatives to the $\Lambda$CDM model. On the other hand, it also allowed us to study the difference between the two BAO datasets that we chose to investigate -- an angular one that is claimed to be cleaned up from the fiducial cosmology ($BAO_\theta$) and a mixed radial and angular ($BAO$). One can see that in terms of DE models, the $BAO_\theta$ dataset leads to significantly less preference for $\Lambda$CDM which we demonstrate with the help of different statistical measures comparing the models. In terms of $\Omega_K$CDM, the two sets of points also possess different behavior, with $BAO_\theta$ preferring a flat universe, while $BAO$ alone having a strong preference for a closed one. We cannot tell to what extend the latter is a numerical artifact due to the fact we marginalize over $H_0 r_d$, which may make it much less numerically sensitive to the curvature of the universe. On the other hand, this approach may offer a way to study how measurements processing affects their bias towards certain models and to allow for a better comparison between different datasets.

\acknowledgments
D.S. is thankful to Bulgarian National Science Fund for support via research grant KP-06-N38/11 

\bibliographystyle{JHEP}
\bibliography{ref}

\providecommand{\href}[2]{#2}\begingroup\raggedright\begin{thebibliography}{10}

\bibitem{Knox:2019rjx}
L.~Knox and M.~Millea, \emph{{Hubble constant hunter\textquoteright{}s guide}},
  \href{https://doi.org/10.1103/PhysRevD.101.043533}{\emph{Phys. Rev. D}
  {\bfseries 101} (2020) 043533}
  [\href{https://arxiv.org/abs/1908.03663}{{\ttfamily 1908.03663}}].

\bibitem{Abdalla:2022yfr}
E.~Abdalla et~al., \emph{{Cosmology intertwined: A review of the particle
  physics, astrophysics, and cosmology associated with the cosmological
  tensions and anomalies}},
  \href{https://doi.org/10.1016/j.jheap.2022.04.002}{\emph{JHEAp} {\bfseries
  34} (2022) 49} [\href{https://arxiv.org/abs/2203.06142}{{\ttfamily
  2203.06142}}].

\bibitem{Aghanim:2018eyx}
{\scshape Planck} collaboration, \emph{{Planck 2018 results. VI. Cosmological
  parameters}},
  \href{https://doi.org/10.1051/0004-6361/201833910}{\emph{Astron. Astrophys.}
  {\bfseries 641} (2020) A6}
  [\href{https://arxiv.org/abs/1807.06209}{{\ttfamily 1807.06209}}].

\bibitem{Riess:2021jrx}
A.G.~Riess et~al., \emph{{A Comprehensive Measurement of the Local Value of the
  Hubble Constant with 1 km s$^{-1}$ Mpc$^{-1}$ Uncertainty from the Hubble
  Space Telescope and the SH0ES Team}},
  \href{https://doi.org/10.3847/2041-8213/ac5c5b}{\emph{Astrophys. J. Lett.}
  {\bfseries 934} (2022) L7}
  [\href{https://arxiv.org/abs/2112.04510}{{\ttfamily 2112.04510}}].

\bibitem{Riess:2022mme}
A.G.~Riess, L.~Breuval, W.~Yuan, S.~Casertano, L.M.~\textasciitilde{}Macri,
  D.~Scolnic et~al., \emph{{Cluster Cepheids with High Precision Gaia
  Parallaxes, Low Zeropoint Uncertainties, and Hubble Space Telescope
  Photometry}},  \href{https://arxiv.org/abs/2208.01045}{{\ttfamily
  2208.01045}}.

\bibitem{Yuan:2022edy}
W.~Yuan, A.G.~Riess, S.~Casertano and L.M.~Macri, \emph{{A First Look at
  Cepheids in a Type Ia Supernova Host with JWST}},
  \href{https://doi.org/10.3847/2041-8213/ac9b27}{\emph{Astrophys. J. Lett.}
  {\bfseries 940} (2022) L17}
  [\href{https://arxiv.org/abs/2209.09101}{{\ttfamily 2209.09101}}].

\bibitem{Poulin:2023lkg}
V.~Poulin, T.L.~Smith and T.~Karwal, \emph{{The Ups and Downs of Early Dark
  Energy solutions to the Hubble tension: a review of models, hints and
  constraints circa 2023}},  \href{https://arxiv.org/abs/2302.09032}{{\ttfamily
  2302.09032}}.

\bibitem{Benisty:2020otr}
D.~Benisty and D.~Staicova, \emph{{Testing late-time cosmic acceleration with
  uncorrelated baryon acoustic oscillation dataset}},
  \href{https://doi.org/10.1051/0004-6361/202039502}{\emph{Astron. Astrophys.}
  {\bfseries 647} (2021) A38}
  [\href{https://arxiv.org/abs/2009.10701}{{\ttfamily 2009.10701}}].

\bibitem{Benisty:2022psx}
D.~Benisty, J.~Mifsud, J.L.~Said and D.~Staicova, \emph{{On the Robustness of
  the Constancy of the Supernova Absolute Magnitude: Non-parametric
  Reconstruction \& Bayesian approaches}},
  \href{https://arxiv.org/abs/2202.04677}{{\ttfamily 2202.04677}}.

\bibitem{Staicova:2021ntm}
D.~Staicova and D.~Benisty, \emph{{Constraining the dark energy models using
  Baryon Acoustic Oscillations: An approach independent of $H_0 \cdot r_d$}},
  \href{https://arxiv.org/abs/2107.14129}{{\ttfamily 2107.14129}}.

\bibitem{Staicova:2022zuh}
D.~Staicova, \emph{{DE Models with Combined H$_{0}$ \textperiodcentered{}
  r$_{d}$ from BAO and CMB Dataset and Friends}},
  \href{https://doi.org/10.3390/universe8120631}{\emph{Universe} {\bfseries 8}
  (2022) 631} [\href{https://arxiv.org/abs/2211.08139}{{\ttfamily
  2211.08139}}].

\bibitem{Chevallier:2000qy}
M.~Chevallier and D.~Polarski, \emph{{Accelerating universes with scaling dark
  matter}}, \href{https://doi.org/10.1142/S0218271801000822}{\emph{Int. J. Mod.
  Phys. D} {\bfseries 10} (2001) 213}
  [\href{https://arxiv.org/abs/gr-qc/0009008}{{\ttfamily gr-qc/0009008}}].

\bibitem{Linder:2002et}
E.V.~Linder, \emph{{Exploring the expansion history of the universe}},
  \href{https://doi.org/10.1103/PhysRevLett.90.091301}{\emph{Phys. Rev. Lett.}
  {\bfseries 90} (2003) 091301}
  [\href{https://arxiv.org/abs/astro-ph/0208512}{{\ttfamily
  astro-ph/0208512}}].

\bibitem{Linder:2005ne}
E.V.~Linder and D.~Huterer, \emph{{How many dark energy parameters?}},
  \href{https://doi.org/10.1103/PhysRevD.72.043509}{\emph{Phys. Rev. D}
  {\bfseries 72} (2005) 043509}
  [\href{https://arxiv.org/abs/astro-ph/0505330}{{\ttfamily
  astro-ph/0505330}}].

\bibitem{Barger:2005sb}
V.~Barger, E.~Guarnaccia and D.~Marfatia, \emph{{Classification of dark energy
  models in the (w(0), w(a)) plane}},
  \href{https://doi.org/10.1016/j.physletb.2006.02.018}{\emph{Phys. Lett. B}
  {\bfseries 635} (2006) 61}
  [\href{https://arxiv.org/abs/hep-ph/0512320}{{\ttfamily hep-ph/0512320}}].

\bibitem{Li:2019yem}
X.~Li and A.~Shafieloo, \emph{{A Simple Phenomenological Emergent Dark Energy
  Model can Resolve the Hubble Tension}},
  \href{https://doi.org/10.3847/2041-8213/ab3e09}{\emph{Astrophys. J. Lett.}
  {\bfseries 883} (2019) L3}
  [\href{https://arxiv.org/abs/1906.08275}{{\ttfamily 1906.08275}}].

\bibitem{Li:2020ybr}
X.~Li and A.~Shafieloo, \emph{{Evidence for Emergent Dark Energy}},
  \href{https://doi.org/10.3847/1538-4357/abb3d0}{\emph{Astrophys. J.}
  {\bfseries 902} (2020) 58}
  [\href{https://arxiv.org/abs/2001.05103}{{\ttfamily 2001.05103}}].

\bibitem{Lazkoz:2005sp}
R.~Lazkoz, S.~Nesseris and L.~Perivolaropoulos, \emph{{Exploring Cosmological
  Expansion Parametrizations with the Gold SnIa Dataset}},
  \href{https://doi.org/10.1088/1475-7516/2005/11/010}{\emph{JCAP} {\bfseries
  11} (2005) 010} [\href{https://arxiv.org/abs/astro-ph/0503230}{{\ttfamily
  astro-ph/0503230}}].

\bibitem{Basilakos:2016nyg}
S.~Basilakos and S.~Nesseris, \emph{{Testing Einstein\textquoteright{}s gravity
  and dark energy with growth of matter perturbations: Indications for new
  physics?}}, \href{https://doi.org/10.1103/PhysRevD.94.123525}{\emph{Phys.
  Rev. D} {\bfseries 94} (2016) 123525}
  [\href{https://arxiv.org/abs/1610.00160}{{\ttfamily 1610.00160}}].

\bibitem{Anagnostopoulos:2017iao}
F.K.~Anagnostopoulos and S.~Basilakos, \emph{{Constraining the dark energy
  models with $H(z)$ data: An approach independent of $H_0$}},
  \href{https://doi.org/10.1103/PhysRevD.97.063503}{\emph{Phys. Rev. D}
  {\bfseries 97} (2018) 063503}
  [\href{https://arxiv.org/abs/1709.02356}{{\ttfamily 1709.02356}}].

\bibitem{Camarena:2021jlr}
D.~Camarena and V.~Marra, \emph{{On the use of the local prior on the absolute
  magnitude of Type Ia supernovae in cosmological inference}},
  \href{https://doi.org/10.1093/mnras/stab1200}{\emph{Mon. Not. Roy. Astron.
  Soc.} {\bfseries 504} (2021) 5164}
  [\href{https://arxiv.org/abs/2101.08641}{{\ttfamily 2101.08641}}].

\bibitem{DiPietro:2002cz}
E.~Di~Pietro and J.-F.~Claeskens, \emph{{Future supernovae data and
  quintessence models}},
  \href{https://doi.org/10.1046/j.1365-8711.2003.06508.x}{\emph{Mon. Not. Roy.
  Astron. Soc.} {\bfseries 341} (2003) 1299}
  [\href{https://arxiv.org/abs/astro-ph/0207332}{{\ttfamily
  astro-ph/0207332}}].

\bibitem{Nesseris:2004wj}
S.~Nesseris and L.~Perivolaropoulos, \emph{{A Comparison of cosmological models
  using recent supernova data}},
  \href{https://doi.org/10.1103/PhysRevD.70.043531}{\emph{Phys. Rev. D}
  {\bfseries 70} (2004) 043531}
  [\href{https://arxiv.org/abs/astro-ph/0401556}{{\ttfamily
  astro-ph/0401556}}].

\bibitem{Perivolaropoulos:2004yr}
L.~Perivolaropoulos, \emph{{Constraints on linear negative potentials in
  quintessence and phantom models from recent supernova data}},
  \href{https://doi.org/10.1103/PhysRevD.71.063503}{\emph{Phys. Rev. D}
  {\bfseries 71} (2005) 063503}
  [\href{https://arxiv.org/abs/astro-ph/0412308}{{\ttfamily
  astro-ph/0412308}}].

\bibitem{Deng:2018jrp}
H.-K.~Deng and H.~Wei, \emph{{Null signal for the cosmic anisotropy in the
  Pantheon supernovae data}},
  \href{https://doi.org/10.1140/epjc/s10052-018-6159-4}{\emph{Eur. Phys. J. C}
  {\bfseries 78} (2018) 755}
  [\href{https://arxiv.org/abs/1806.02773}{{\ttfamily 1806.02773}}].

\bibitem{Nunes:2020hzy}
R.C.~Nunes, S.K.~Yadav, J.F.~Jesus and A.~Bernui, \emph{{Cosmological parameter
  analyses using transversal BAO data}},
  \href{https://doi.org/10.1093/mnras/staa2036}{\emph{Mon. Not. Roy. Astron.
  Soc.} {\bfseries 497} (2020) 2133}
  [\href{https://arxiv.org/abs/2002.09293}{{\ttfamily 2002.09293}}].

\bibitem{Pan-STARRS1:2017jku}
{\scshape Pan-STARRS1} collaboration, \emph{{The Complete Light-curve Sample of
  Spectroscopically Confirmed SNe Ia from Pan-STARRS1 and Cosmological
  Constraints from the Combined Pantheon Sample}},
  \href{https://doi.org/10.3847/1538-4357/aab9bb}{\emph{Astrophys. J.}
  {\bfseries 859} (2018) 101}
  [\href{https://arxiv.org/abs/1710.00845}{{\ttfamily 1710.00845}}].

\bibitem{Handley:2015fda}
W.J.~Handley, M.P.~Hobson and A.N.~Lasenby, \emph{{PolyChord: nested sampling
  for cosmology}}, \href{https://doi.org/10.1093/mnrasl/slv047}{\emph{Mon. Not.
  Roy. Astron. Soc.} {\bfseries 450} (2015) L61}
  [\href{https://arxiv.org/abs/1502.01856}{{\ttfamily 1502.01856}}].

\bibitem{Lewis:2019xzd}
A.~Lewis, \emph{{GetDist: a Python package for analysing Monte Carlo samples}},
   \href{https://arxiv.org/abs/1910.13970}{{\ttfamily 1910.13970}}.

\bibitem{eBOSS:2020yzd}
{\scshape eBOSS} collaboration, \emph{{Completed SDSS-IV extended Baryon
  Oscillation Spectroscopic Survey: Cosmological implications from two decades
  of spectroscopic surveys at the Apache Point Observatory}},
  \href{https://doi.org/10.1103/PhysRevD.103.083533}{\emph{Phys. Rev. D}
  {\bfseries 103} (2021) 083533}
  [\href{https://arxiv.org/abs/2007.08991}{{\ttfamily 2007.08991}}].

\bibitem{Liddle:2007fy}
A.R.~Liddle, \emph{{Information criteria for astrophysical model selection}},
  \href{https://doi.org/10.1111/j.1745-3933.2007.00306.x}{\emph{Mon. Not. Roy.
  Astron. Soc.} {\bfseries 377} (2007) L74}
  [\href{https://arxiv.org/abs/astro-ph/0701113}{{\ttfamily
  astro-ph/0701113}}].

\bibitem{Jeffreys:1939xee}
H.~Jeffreys, \emph{{The Theory of Probability}}, Oxford Classic Texts in the
  Physical Sciences (1939).

\bibitem{Bernardo:2022pyz}
R.C.~Bernardo, D.~Grand\'on, J.~Levi~Said and V.H.~C\'ardenas, \emph{{Dark
  energy by natural evolution: Constraining dark energy using Approximate
  Bayesian Computation}},  \href{https://arxiv.org/abs/2211.05482}{{\ttfamily
  2211.05482}}.

\bibitem{Mehrabi:2022mnr}
A.~Mehrabi and J.~Levi~Said, \emph{{Gaussian discriminators between $\varLambda
  $CDM and wCDM cosmologies using expansion data}},
  \href{https://doi.org/10.1140/epjc/s10052-022-10737-8}{\emph{Eur. Phys. J. C}
  {\bfseries 82} (2022) 806}
  [\href{https://arxiv.org/abs/2203.01817}{{\ttfamily 2203.01817}}].

\bibitem{Ryan:2018aif}
J.~Ryan, S.~Doshi and B.~Ratra, \emph{{Constraints on dark energy dynamics and
  spatial curvature from Hubble parameter and baryon acoustic oscillation
  data}}, \href{https://doi.org/10.1093/mnras/sty1922}{\emph{Mon. Not. Roy.
  Astron. Soc.} {\bfseries 480} (2018) 759}
  [\href{https://arxiv.org/abs/1805.06408}{{\ttfamily 1805.06408}}].

\bibitem{Vagnozzi:2020dfn}
S.~Vagnozzi, A.~Loeb and M.~Moresco, \emph{{Eppur è piatto? The Cosmic
  Chronometers Take on Spatial Curvature and Cosmic Concordance}},
  \href{https://doi.org/10.3847/1538-4357/abd4df}{\emph{Astrophys. J.}
  {\bfseries 908} (2021) 84}
  [\href{https://arxiv.org/abs/2011.11645}{{\ttfamily 2011.11645}}].

\bibitem{DiValentino:2019qzk}
E.~Di~Valentino, A.~Melchiorri and J.~Silk, \emph{{Planck evidence for a closed
  Universe and a possible crisis for cosmology}},
  \href{https://doi.org/10.1038/s41550-019-0906-9}{\emph{Nature Astron.}
  {\bfseries 4} (2019) 196} [\href{https://arxiv.org/abs/1911.02087}{{\ttfamily
  1911.02087}}].

\bibitem{Vagnozzi:2020rcz}
S.~Vagnozzi, E.~Di~Valentino, S.~Gariazzo, A.~Melchiorri, O.~Mena and J.~Silk,
  \emph{{The galaxy power spectrum take on spatial curvature and cosmic
  concordance}}, \href{https://doi.org/10.1016/j.dark.2021.100851}{\emph{Phys.
  Dark Univ.} {\bfseries 33} (2021) 100851}
  [\href{https://arxiv.org/abs/2010.02230}{{\ttfamily 2010.02230}}].

\bibitem{Yang:2022kho}
W.~Yang, W.~Giar\`e, S.~Pan, E.~Di~Valentino, A.~Melchiorri and J.~Silk,
  \emph{{Revealing the effects of curvature on the cosmological models}},
  \href{https://doi.org/10.1103/PhysRevD.107.063509}{\emph{Phys. Rev. D}
  {\bfseries 107} (2023) 063509}
  [\href{https://arxiv.org/abs/2210.09865}{{\ttfamily 2210.09865}}].

\bibitem{Jimenez:2022asc}
R.~Jimenez, A.R.~Khalife, D.F.~Litim, S.~Matarrese and B.D.~Wandelt, \emph{{Why
  is zero spatial curvature special?}},
  \href{https://arxiv.org/abs/2210.10102}{{\ttfamily 2210.10102}}.

\bibitem{Wei:2016xti}
J.-J.~Wei and X.-F.~Wu, \emph{{An Improved Method to Measure the Cosmic
  Curvature}}, \href{https://doi.org/10.3847/1538-4357/aa674b}{\emph{Astrophys.
  J.} {\bfseries 838} (2017) 160}
  [\href{https://arxiv.org/abs/1611.00904}{{\ttfamily 1611.00904}}].

\bibitem{Stevens:2022evv}
J.~Stevens, H.~Khoraminezhad and S.~Saito, \emph{{Constraining the spatial
  curvature with cosmic expansion history in a cosmological model with a
  non-standard sound horizon}},
  \href{https://arxiv.org/abs/2212.09804}{{\ttfamily 2212.09804}}.

\end{thebibliography}\endgroup
\appendix
\section{Appendix}
\begin{figure}[!h]
        \centering
\includegraphics[width=0.41\textwidth]{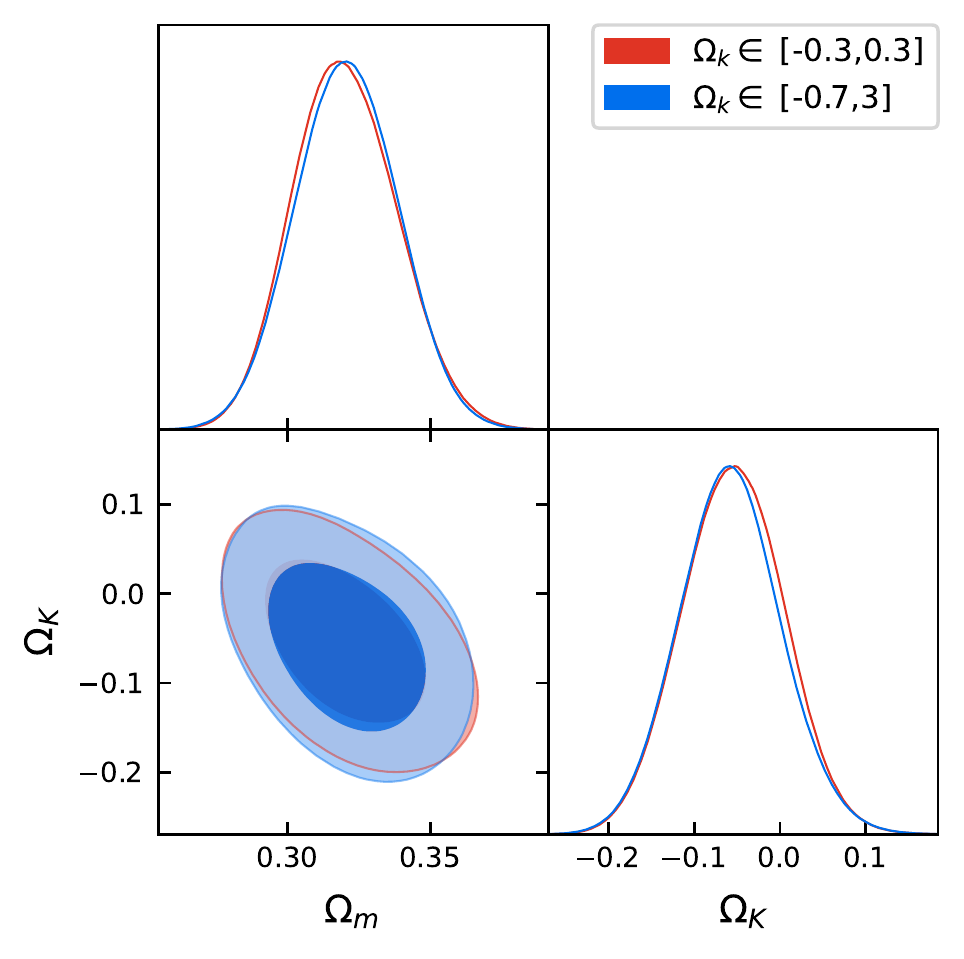}
\includegraphics[width=0.41\textwidth]{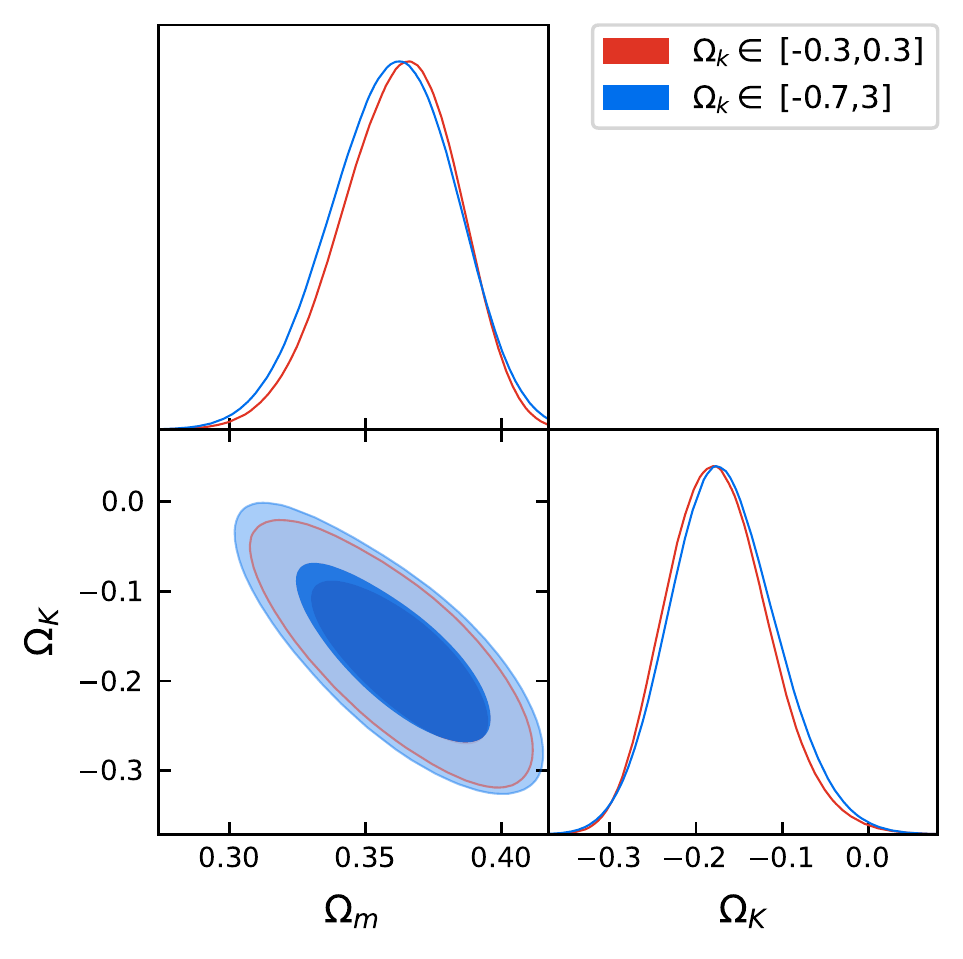}
\caption{\it{Posterior distribution for $\Omega_m$ and $\Omega_K$ for BAO (left panel) and BAO+SN (right panel) datasets}}
        \label{fig:Ok_BAO}
\end{figure}

\begin{figure}[!h]
        \centering
\includegraphics[width=0.41\textwidth]{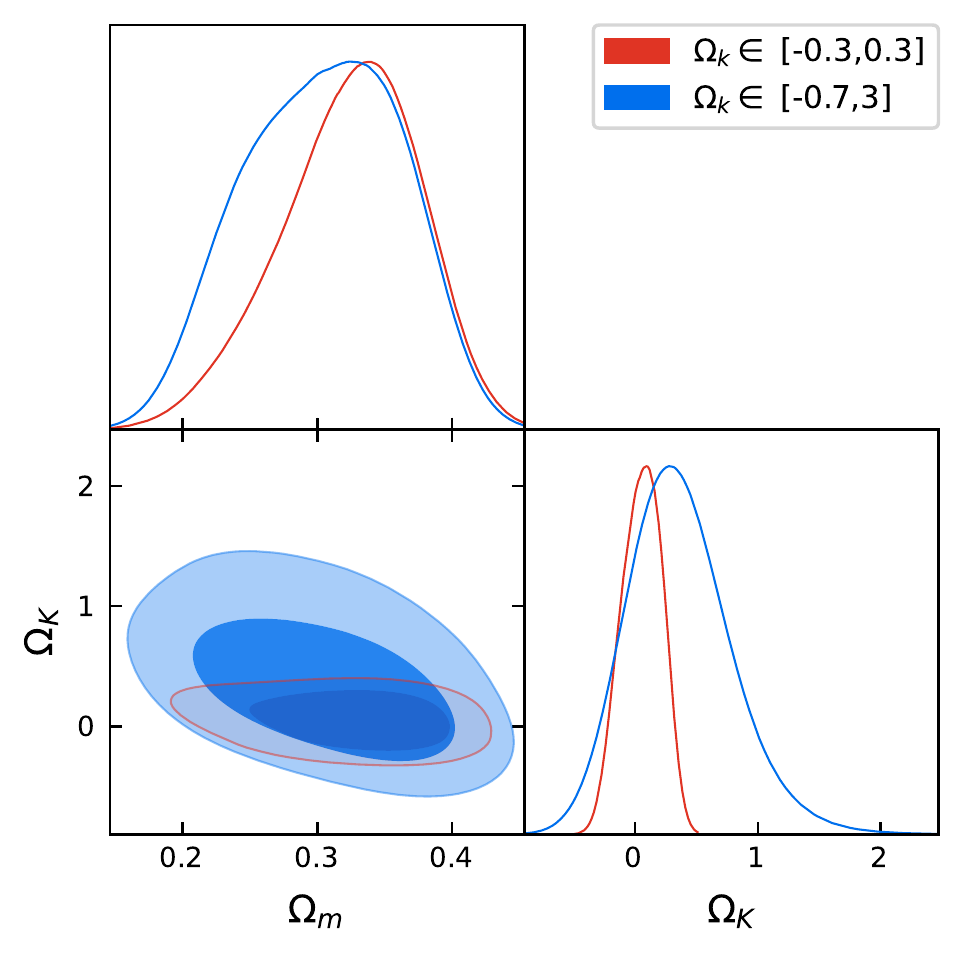}
\includegraphics[width=0.41\textwidth]{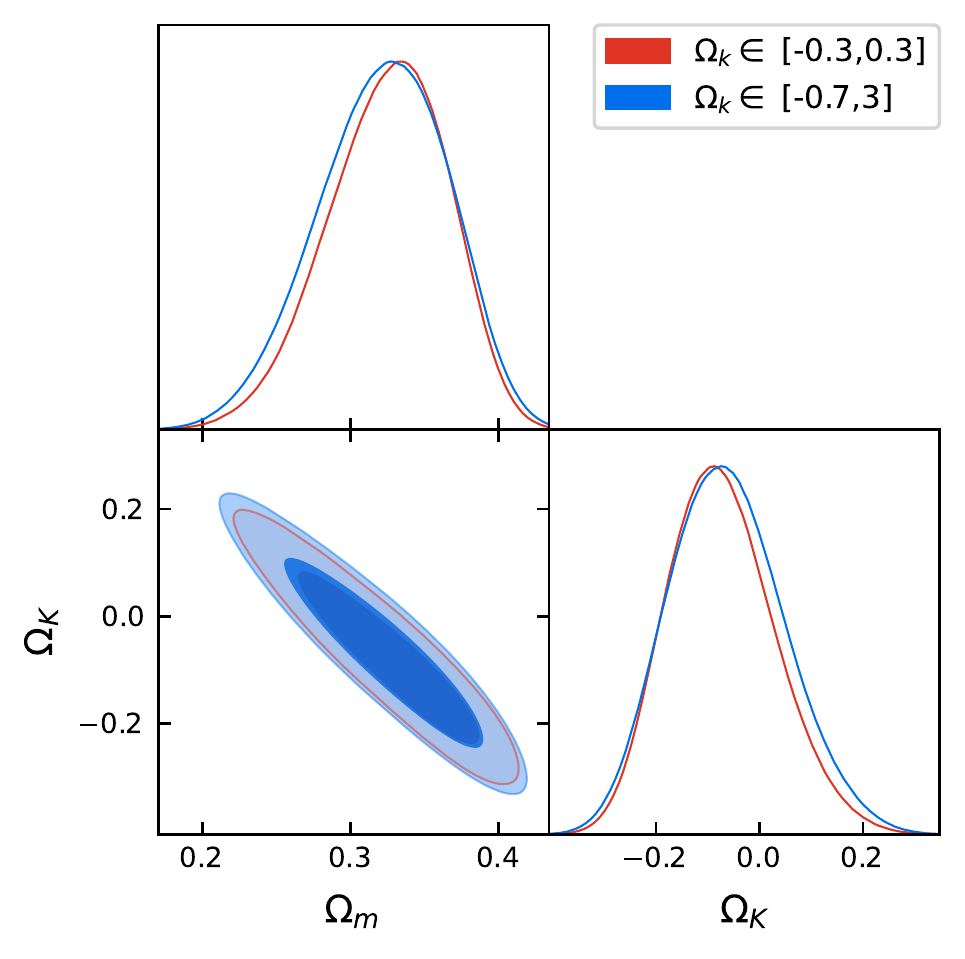}
\caption{\it{Posterior distribution for $\Omega_m$ and $\Omega_K$ for $BAO_\theta$ (left) and $BAO_\theta$+SN (right) datasets}}
        \label{fig:Ok_BAOTheta}
\end{figure}

\end{document}